# Lead-Related Quantum Emitters in Diamond

*Matthew E. Trusheim*[1,*,†], *Noel H. Wan*[1,*], *Kevin C. Chen*[1,*], *Christopher J. Ciccarino*[2,3], *Ravishankar Sundararaman*[4], *Girish Malladi*[5], *Eric Bersin*[1], *Michael Walsh*[1], *Benjamin Lienhard*[1], *Hassaram Bakhru*[5], *Prineha Narang*[2], *and Dirk Englund*[1]

**Affiliations**

[1]Department of Electrical Engineering and Computer Science, Massachusetts Institute of Technology, Cambridge, Massachusetts 02139, USA
[2]John A. Paulson School of Engineering and Applied Sciences, Harvard University, Cambridge, Massachusetts 02138, USA
[3]Department of Chemistry and Chemical Biology, Harvard University, Cambridge, Massachusetts, 02138, USA
[4]Department of Materials Science and Engineering, Rensselaer Polytechnic Institute, Troy, NY, United States
[5]College of Nanoscale Science and Engineering, Suny Poly, 257 Fuller Road, Albany, NY 12203, United States

[†]mtrush@mit.edu

[*]These authors contributed equally to this work

**Abstract**

We report on quantum emission from Pb-related color centers in diamond following ion implantation and high temperature vacuum annealing. First-principles calculations predict a negatively-charged Pb-vacancy center in a split-vacancy configuration, with a zero-phonon transition around 2.3 eV. Cryogenic photoluminescence measurements performed on emitters in nanofabricated pillars reveal several transitions, including a prominent doublet near 520 nm. The splitting of this doublet, 2 THz, exceeds that reported for other group-IV centers. These observations are consistent with the PbV center, which is expected to have the combination of narrow optical transitions and stable spin states, making it a promising system for quantum network nodes.

**Introduction**

Quantum emitters in diamond and other wide-bandgap materials are promising systems for quantum information processing[1,2]. In particular, the nitrogen-vacancy center in diamond (NV) has been shown to possess both long-lived spin states[3] and a high-fidelity spin-photon interface[4], enabling quantum networking protocols[5]. However, the NV's optical properties remain a challenge: only a small fraction of the fluorescence branches into the coherent zero-phonon line (ZPL), as given by a Debye-Waller factor (DWF) of ~0.03[6], and spectral diffusion broadens the ZPL significantly in nanophotonic devices[7,8,9].

These limitations have spurred a search for alternative quantum emitters. These include the negatively-charged silicon-vacancy (SiV⁻) center, which has a larger DWF ~0.8[10] and stable optical transitions even in nanophotonic structures[11] due to its crystallographic inversion symmetry[12,13]. To achieve long spin coherence in SiV⁻ centers, it is necessary to cool the sample such that $k_B T \ll h\Delta_{GS}$, where $\Delta_{GS}$ is the ground state orbital splitting, to reduce phonon absorption in the ground state manifold. For the SiV⁻ defect, with $\Delta_{GS}$ = 50 GHz, temperatures below 500 mK are required to achieve millisecond coherence times[14]. Heavier group IV-vacancy centers are expected to share the favorable inversion symmetry, yielding narrow optical lines, but with larger ground state orbital splittings that could increase spin coherence times[15]. In previous reports of group IV defects, the germanium-vacancy has been shown to have narrow optical lines[16,17] and $\Delta_{GS}$ = 150 GHz, and the tin-vacancy has been observed[18,19] with $\Delta_{GS}$ = 850 GHz. The increase in $\Delta_{GS}$ effectively relaxes the temperature requirement for achieving long coherence times, motivating the search for defects with heavier elements.

Here we report the observation of color centers associated with the heaviest naturally-occurring group IV element, lead (Pb). Our first-principles density functional theory (DFT) calculations predict a stable negatively-charged Pb-vacancy (PbV) color center in the diamond lattice. We created Pb-related color centers through an ion implantation and annealing process and characterized them through photoluminescence spectroscopy at cryogenic temperatures. As discussed below, we find good agreement between DFT predictions of the transition energies and our spectroscopy.

**First Principles Calculations of Defects and Spectra**

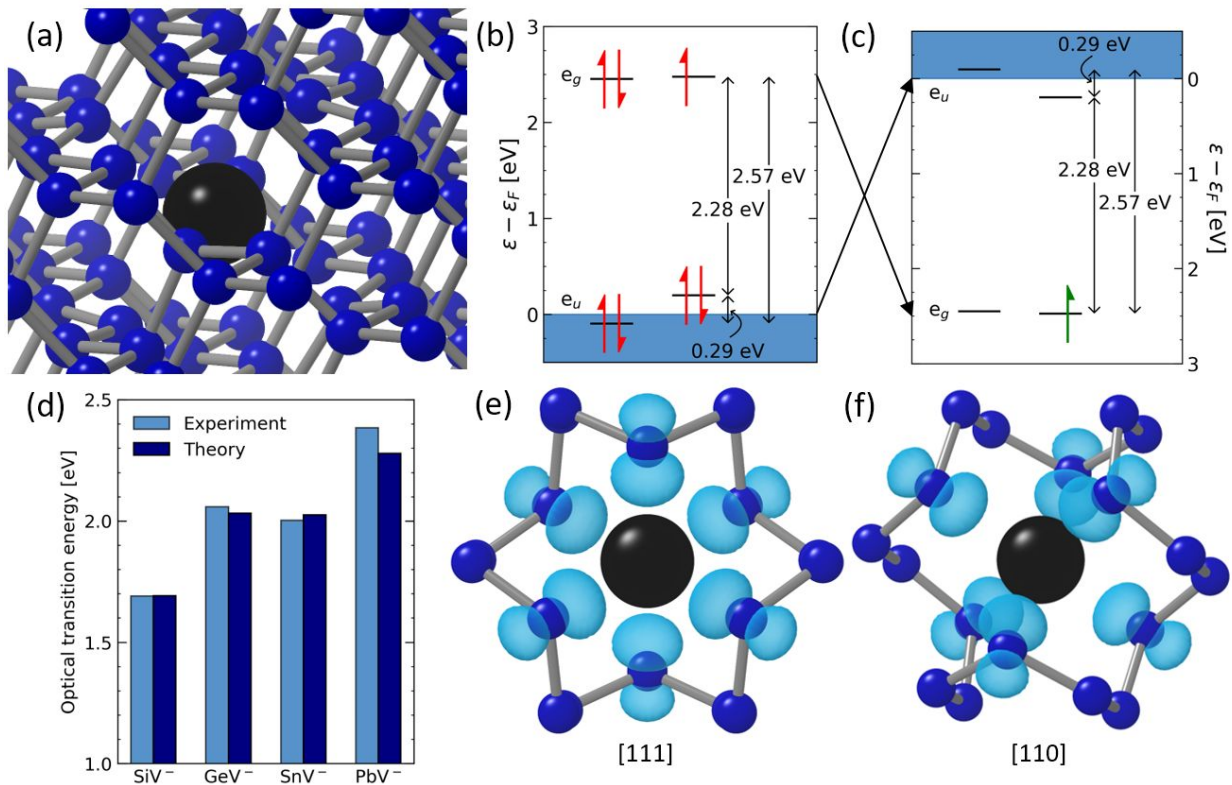

*Figure 1: PbV calculations from first principles. (a) A Pb atom (black) in a split-vacancy configuration within the diamond lattice (blue spheres: carbon atoms). (b) The electron orbital structure of the PbV defect in a negative charge state. Both the $e_u$ and $e_g$ orbitals are split as a result of strong spin-orbit coupling. (c) The same PbV orbital structure from a hole perspective. (d) Comparison of predicted optical transition energies of group-IV defects with experimental data[18,20,21] and present work. (e,f) Electron density associated with the negatively-charged PbV viewed from the [111] and [110] axes, respectively. The density is primarily distributed to the six nearby carbon atoms of the diamond lattice.*

We performed first-principles simulations of the PbV defect (platform JDFTx[22]) using a 512 atom supercell which is modeled using fully-relativistic ultrasoft pseudopotentials[23] and a Perdew-Burke-Ernzerhof (PBE) exchange-correlation functional. After fully relaxing the system using gamma point integration, the defect is found to be stable in the split-vacancy configuration (Figure 1a). We determine the stable charge state to be negatively charged by using a novel charge correction scheme[24]. The predicted orbital structure, which is composed of two pairs of orbitals, $e_u$ and $e_g$, is shown in Figure 1b. These orbitals are both nondegenerate due to spin-orbit coupling effects, which have been key contributors to the observed spectra of lighter group-IV defects[17,25].

The seven-electron orbital configuration of the PbV is most simply described in terms of a single hole picture. This is schematically presented in Figure 1c, where we follow convention and flip the energy axis. From this perspective, the potential transitions of the hole from excited states ($e_u$ orbitals) to ground states ($e_g$ orbitals) correspond to optical emission at different energies. The difference in energy between $e_g$ orbitals is the ground-state splitting $\Delta_{GS}$, while the splitting between $e_u$ orbitals is the excited state splitting $\Delta_{ES}$, consistent with previous convention[25]. We determine these energies by including spin-orbit coupling in our calculations. Both the ground and excited state splittings increase with the size of the group-IV atom, with the PbV having the largest splittings $\Delta_{GS}$= 25.4 meV, $\Delta_{ES}$= 292 meV. Note that our calculations do not include additional factors which would affect the observed orbital splittings, such as the Ham reduction factor [26]; including this contribution would lower the magnitude of the predicted $\Delta_{GS}$, $\Delta_{ES}$.

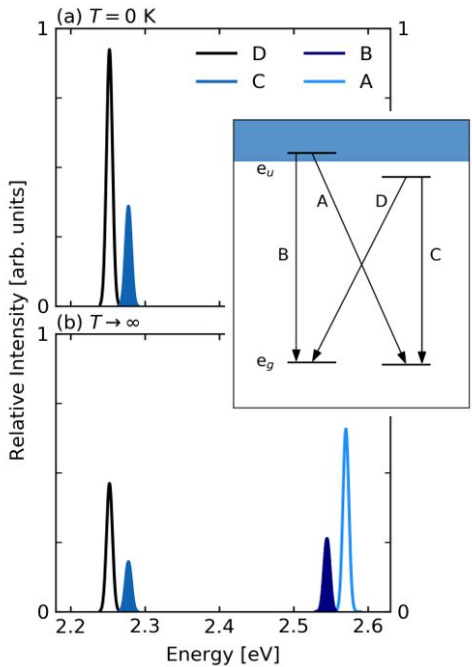

*Figure 2: Calculated relative emission intensity from first-principles. Dipole matrix elements between each PbV orbital are calculated and compared for different temperatures. (a) Predicted emission at zero temperature. The emitting transitions originate from the lower-energy $e_u$ orbital (i.e., channels D and C), under the assumption that thermal equilibrium is achieved prior to emission[27]. In the high temperature limit presented in panel (b), transitions to all four orbitals are thermodynamically allowed, resulting in a four-peak structure. Filled peaks indicate emission polarized along the PbV [111] direction (transitions B, C), while unfilled peaks are polarized perpendicular to the defect axis.*

To predict the strengths of the optical transitions via Fermi's Golden rule, we calculate the momentum matrix elements $\langle \psi_n | \vec{p} | \psi_{n'} \rangle$ for PbV states n and n'. Figure 2 presents the relative coupling strengths of the four zero-phonon transitions A, B, C, and D at T = 0 and T → ∞. Electron-phonon interactions[28,29] would introduce additional spectral features and broaden the linewidths of each transition. There are only two peaks at low temperatures because we assume that the $e_u$ orbitals reach thermal equilibrium prior to emission[27]. Thus, de-excitation of an excited hole would only occur out of the lower-energy $e_u$ orbital, which is located within the band gap. To model emission at elevated temperatures, we introduce a Boltzmann distribution, which in the high temperature limit allows all four orbitals to be equally occupied. As a result, four peaks are expected in the emission spectrum (Figure 2b). From these calculations, we also obtain the emission polarization. As shown, two of the four transitions (B and C) are polarized along the PbV [111] axis. The two other emission channels (A and D) have polarization perpendicular to the defect axis. This polarization trend was observed for each of the lighter group-IV defects as well.

**Experiment**

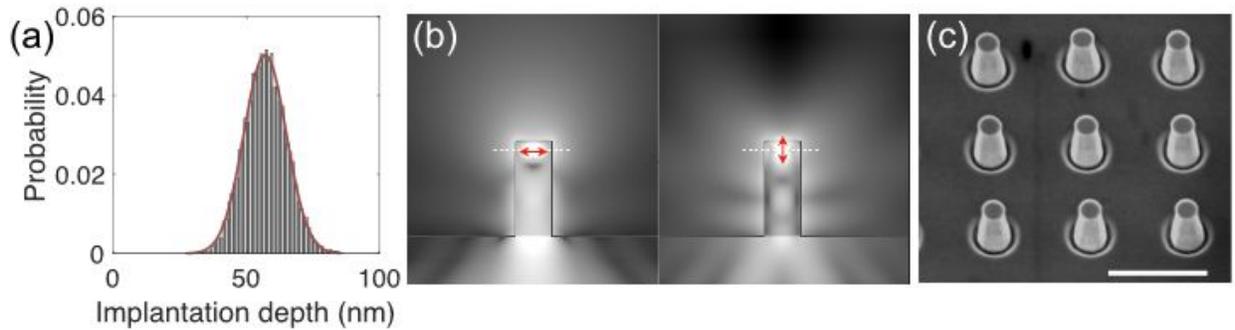

Figure 3: Sample preparation. (a) Simulated Pb ion probability distribution, centered around (58 ±8) nm in depth. (b) Field profile of an emitter in a nanopillar with a diameter of 225 nm and height of 600 nm. Red arrow: polarization of a dipole emitter. Dashed white line: predicted depth of Pb-related emitters (c) Scanning electron micrograph of the fabricated nanopillar array. Scale bar: 1 µm.

We prepared the sample by ion implantation into commercially-available type-IIa diamond (< 5 ppb [N],[B]; Element6) at an energy of 350 keV and a dose of $10^9$ Pb cm$^{-2}$ (Varian Extrion 10 - 400 keV ion implanter). Stopping range of ions in matter (SRIM) calculations[30] (Figure 3a) predict that this implantation produces a Pb layer with a mean depth of (58 ± 8) nm. Each implanted Pb ion is predicted to produce ~ 2000 vacancies during implantation; although an order of magnitude more than expected for other species such as nitrogen, the predicted peak vacancy density of ~$10^{17}$ cm$^{-3}$ is still well below the graphitization damage threshold for diamond[31].

Following implantation, we annealed the sample under high vacuum (< $10^{-7}$ mbar) at 1200 °C for two hours and cleaned it in boiling acid (see Supplemental). The resulting sample had a density of emitters too high to spatially isolate a single color center (see Supplemental). Given our measurement spot size of ~ 200 nm, this suggests that creation yield of Pb-related emitters (density > 1 per 200 x 200 $nm^2$), from implanted Pb ions (10 per 1 x 1 $um^2$), approaches unity.

To isolate single Pb-related emitters, we fabricated nanopillars into the diamond using a combination of electron beam lithography and reactive ion etching (Figure 3b,c, see Supplemental for details). The pillar diameters varied from 150 to 325 nm, with a height of 670 nm. For collection using an NA = 0.9 objective, finite-difference time-domain (FDTD) calculations indicate that the pillars additionally provide a 5 to 10X fluorescence collection efficiency enhancement for an emitter located at a depth of 60 nm as compared to that in an unpatterned diamond. The scanning electron micrograph in Figure 3c shows the set of nanopillars with a diameter of 225 nm that are primarily investigated in this work.

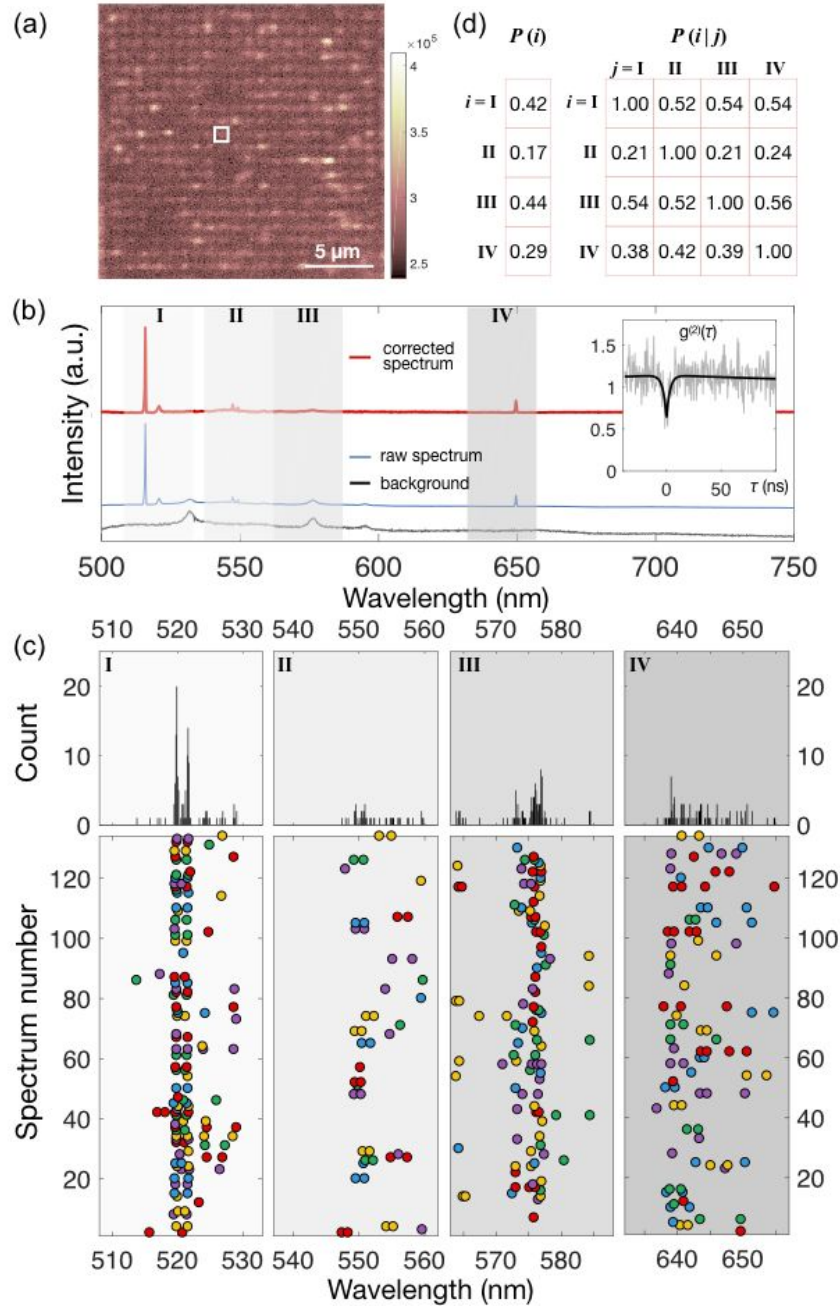

Figure 4: Cryogenic (4 K) emitter characterization. (a) Confocal scan of Pb-related emitters in nanopillars. Colorbar: counts per second. White box indicates a single emitter of interest as described below. (b) Emission spectrum of boxed emitter. Four regions with observed emission lines are shaded and labeled. Inset: antibunched emission with $g^2(0) = 0.52$ ($g^2(0) = 0.28$ after background correction) and excited state lifetime of 3.0 ns. (c) Summary of photoluminescence spectra from 133 pillars. Circles indicate the center of an observed emission peak, and colors correspond to individual pillars horizontally across spectral regions. Top: histogram of emission peak locations. (d) Overall and conditional probabilities for the observation of emission in each region.

Figure 4 summarizes fluorescence spectroscopy of Pb-related emitters in these nanopillars, acquired using a cryogenic confocal microscope (NA = 0.9) at a temperature of 4 K. Figure 4a shows a representative confocal scan of a 20 x 20 array of 225 nm diameter pillars using an excitation laser at 450 nm (750 µW). Bright spots corresponding to emitters in individual nanopillars are clearly visible. Figure 4b shows a representative spectrum from the pillar indicated by the white box in Figure 4a. This pillar contains a single emitter (red curve) as verified by second-order autocorrelation measurements (see inset of 4b). After correcting for the observed sample-independent background (gray), we measure $g^{(2)}(0)$ = 0.28, which indicates quantum emission from a single emitter. The measured antibunching timescale gives a lifetime of $T_1$ = 3 ns. Additional spectra from the 225 nm diameter posts are displayed in the Supplemental Materials.

Figure 4c shows a statistical investigation of Pb-related emitter spectral properties. We analyze spectra in the four distinct regions indicated as I-IV in Figure 4b. Figure 4c summarizes the fluorescence emission from the 133 brightest pillars. We fit the location of emission peaks observed for each nanopillar in the dataset, and plot the peak locations as circles in Figure 4c. We then bin the peak locations to create an inhomogeneous spectrum of Pb-related emitters (Figure 4c, top). Finally, we calculate the overall ($P(i)$) and conditional ($P(i \mid j)$) probabilities of observing an emission line in a given region (Figure 4d). We do not observe any significant correlation between emission regions, with $P(i) \approx P(i|j) \ \forall \ i \neq j$.

We tentatively attribute the lines in Regions III and IV (575 and 640 nm) to neutral and negatively charged NV centers[32], respectively, which are formed in the implantation process from residual nitrogen atoms naturally present in the diamond. We note that many of these emitters have large shifts from the unstrained NV ZPL position, which could be due to large strains induced by Pb implantation. The presence of a significant number of peaks in both Region III ($NV^0$) and Region IV ($NV^-$) indicates that the charge environment of the implanted layer is not uniform. In addition, the lack of correlation between the regions shows that the NVs that do exist are preferentially in a single charge state, unlike reported photochromic NV centers[33,34] in similar ultrapure type-IIa diamond without Pb implant.

The emission lines in Region I have not been previously reported, and we attribute them to Pb-related defect centers. The observed inhomogeneous linewidths (0.355 nm and 0.323 nm) of the prominent doublet at 520 nm in Region I are broadened between sites. However, individual diffraction-limited spots do display lines narrower than our spectrometer-limited resolution of 0.1 nm. The measured splitting of the Region I peaks, 1.8 nm (2 THz), is greater than that measured for the ground state of the $SiV^{25}$, $GeV^{17}$, and $SnV^{18}$.

**Discussion**
The theoretically calculated and experimentally measured PbV-emitter ZPL are in relatively good agreement, with the experimentally measured ZPL doublet at 520 nm close to the simulated 544 nm. Theory predicts a four-level state structure which should point towards split peaks in the emission spectra. At cryogenic temperatures, we find that the ground-state splitting is 2 THz, which is relatively close to the theoretical predicted value of 6.04 THz. The discrepancy between these values is likely due to the theoretical value being too large (see Section 2). However, emission associated with the upper excited state, which is expected to be present at higher temperatures, was not conclusively observed in experiment. These transitions could be outside of the accessible spectral range (> 500 nm) in our measurement setup.

Nonetheless, the line featured at 520 nm is prominent and has yet to be reported in diamond. For this reason, we assign this peak to the PbV color center.

In addition to this pronounced line, cryogenic spectroscopy revealed other spectral features. There are several possible causes for these additional lines. We note that SnV centers produced by implantation show a broad (> 30 nm) inhomogeneous distribution[18,19] following annealing at temperatures below 2000 °C, including several peaks attributed to intermediate defect states. These states are eliminated upon high-pressure high-temperature annealing[18], which has been shown to reduce strain in the diamond lattice in addition to altering the thermodynamic condition for defect stability. Pb implantation produces more vacancies per ion than Sn and within a smaller volume, and thus can significantly alter the lattice during implantation in a similar fashion. Intermediate defect states and highly strained local environments are therefore candidates for the observed emission lines.

The charge stability of the PbV defect is a another potential cause of additional spectral features. We theoretically predict a charge transition level to be 2.71 eV above the valence band maximum. This suggests that photoionization could contribute to spectra by illumination into the phonon sideband under our pump at 450 nm (2.76 eV), as seen with NV[35,36] and SiV centers[37]. We observed fluorescence instability in some centers under blue illumination (Supplemental Materials), which could indicate the presence of an alternate charge state. Additionally, other spectral features are observed in fluorescence under 532 nm illumination (appearance of lines around 590 and 715 nm, see Supplement). Future measurements, such as photoluminescence excitation spectroscopy under varying electromagnetic and strain fields, will be required to fully determine the optical properties and electronic structure of the observed Pb-related emitters.

In conclusion, we have shown quantum emission related to Pb defects in diamond for the first time. We observe antibunched emission near 520 nm, which agrees with first-principle calculations predicting a stable, negatively-charged split-vacancy PbV center. Importantly, we measure a splitting of 2 THz which exceeds the ground state splitting of the SiV, GeV, and SnV centers, in agreement with our simulations for PbV. This could enable long spin coherence at elevated temperatures, which is of central importance for quantum memories in proposed quantum networks and modular quantum computing schemes.

## Supplemental Materials

**Cryogenic Measurement Setup Description**

Cryogenic experiments are performed using a home-built scanning confocal microscope. The samples are cooled using a closed-cycle helium cryostat (Montana Instruments) and are imaged through a 0.9 NA vacuum objective. Piezo stages (Attocube) inside the chamber allow micrometer positioning of the sample, and galvanometer mirrors (Nutfield QS-5 OPD) allow nanometer scanning of the confocal spot. For excitation, 532 nm light is generated by a Coherent Verdi G5 laser and 450 nm light by a laser diode. Collected fluorescence is routed to either an avalanche photodetector (Excelitas SPCM-AQRH-14) for count-rate measurements, a free-space spectrometer (Princeton Instruments, IsoPlane SCT 320) for spectral characterization, or a Hanbury Brown and Twiss interferometer consisting of a fiber beamsplitter (Evanescent Optics) and two fiber-input APDs (Excelitas SPCM-AQRH-14-FC and Perkin Elmer SPCM-AQRH-14-FC) for second-order correlation measurements using a picosecond resolution time tagger (Pico-Quant PicoHarp 300). For measurements using blue light, a 450 nm bandpass filter and 500 nm shortpass filter (Thorlabs) are used on the excitation laser path, while a 500 nm longpass filter (Thorlabs) is used on the emission. For second-order correlation measurements, a tunable shortpass filter (Semrock) was used to exclude emission from Region III and IV. A 532 nm bandpass filter on the excitation path and 532 nm notch filter (Thorlabs) were used in experiments employing 532 nm excitation.

**Unpatterned Implant Characterization**

Without the fabrication of nanoposts, the Pb-implanted diamond is bright, without spatially separable emitters (Figure S1a). A representative second-order correlation measurement taken on the unpatterned, Pb-implanted diamond is shown in Figure S1b. A small antibunching dip is observed, $g^{(2)}(0) = 0.97$, indicating a very low, but non-zero, fraction of correlated single-photon emission.

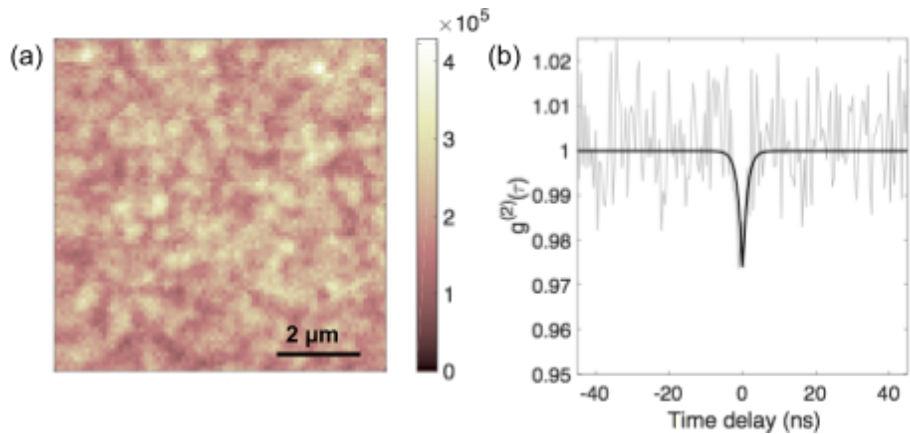

*Figure S1: Unpatterned implant characterization. (a) Confocal scan. Colorbar: counts per second. (b) Representative second-order correlation taken on an unpatterned Pb-implanted region.*

**Nanopillar Fabrication**

We fabricated nanopillars in the implanted diamond using a dry reactive-ion etching (RIE) process[38,39]. A 180 nm-thick silicon nitride (SiN) serves as our hard mask, which is deposited using plasma-enhanced chemical vapor deposition. We patterned the SiN hard mask with ZEP-520A electron beam resist and tetrafluoromethane dry etching after which we etched the diamond using an inductively-coupled oxygen plasma RIE. We then removed the mask using hydrofluoric acid and cleaned the diamond in a boiling reflux of 1:1:1 hydrochloric, sulfuric and perchloric acids. Finally, we annealed the diamond at 1200°C as described in the text and cleaned the diamond using a similar mixture of boiling acids.

**Individual Spectra under 450 nm Illumination**

Figure S2 shows additional representative spectra taken under 450 nm illumination. While lines in each of the four regions I-IV shown are present in individual spectra, no consistent correlation is observed between them.

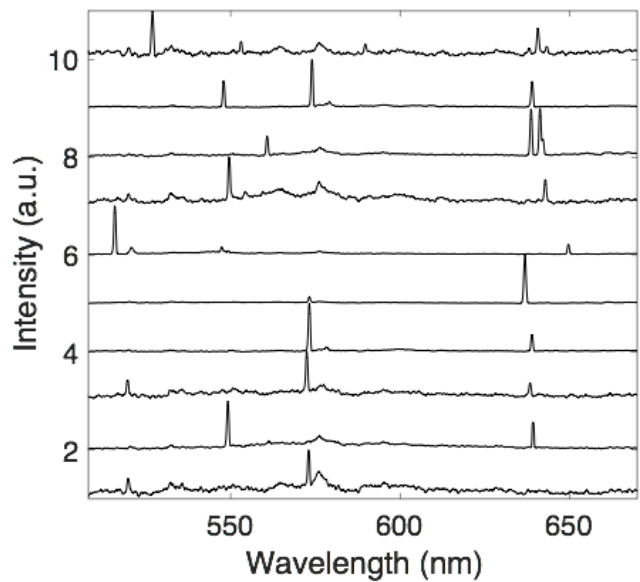

*Figure S2: Example spectra under 450 nm excitation. Lines in regions I-IV are observed, without consistent correlation.*

**Pb Emitters under 532 nm illumination**

The 520 nm Pb-related ZPL observed under blue excitation should not be excited under 532 nm illumination, as the energy is too low. However, we observe two distinct emission lines under 532 nm illumination that are not present under 450 nm (Figure S3): 715 and 590 nm. We additionally observe a wide spread in emission lines from 640 to 670 nm, which we tentatively attribute to NV centers under extreme strain. As 1 GPa external stress results in an NV level shift of ~ 1 THz, a 30 nm (~ 20 THz) shift implies an internal strain of 0.02, taking the Young's modulus of diamond to be ~ 1 TPa/strain. This implies that Pb implantation significantly disrupts the diamond lattice on the nanoscale. A representative selection of spectra under 532 nm excitation are shown in Figure S2: no correlation between emission at 590 and 715 nm is seen.

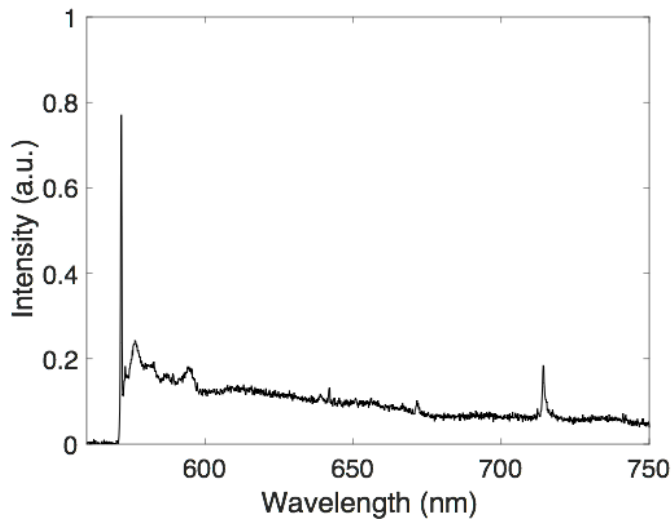

Figure S3: Average inhomogeneous PbV spectrum under 532 nm illumination at 4 K. Mean of emission intensity from 205 nanoposts. 572 nm, diamond Raman line.

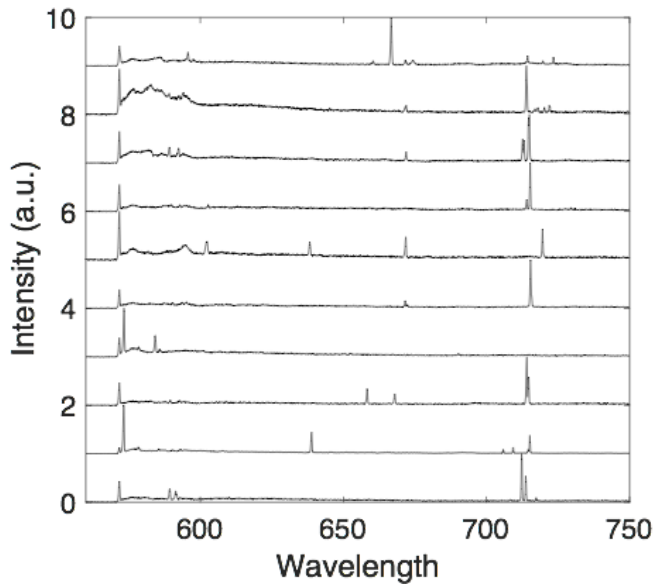

Figure S4: Example Pb emitter spectra under 532 nm illumination. NV-related lines are widely dispersed; peaks at 590 and 715 nm are visible, but independent.

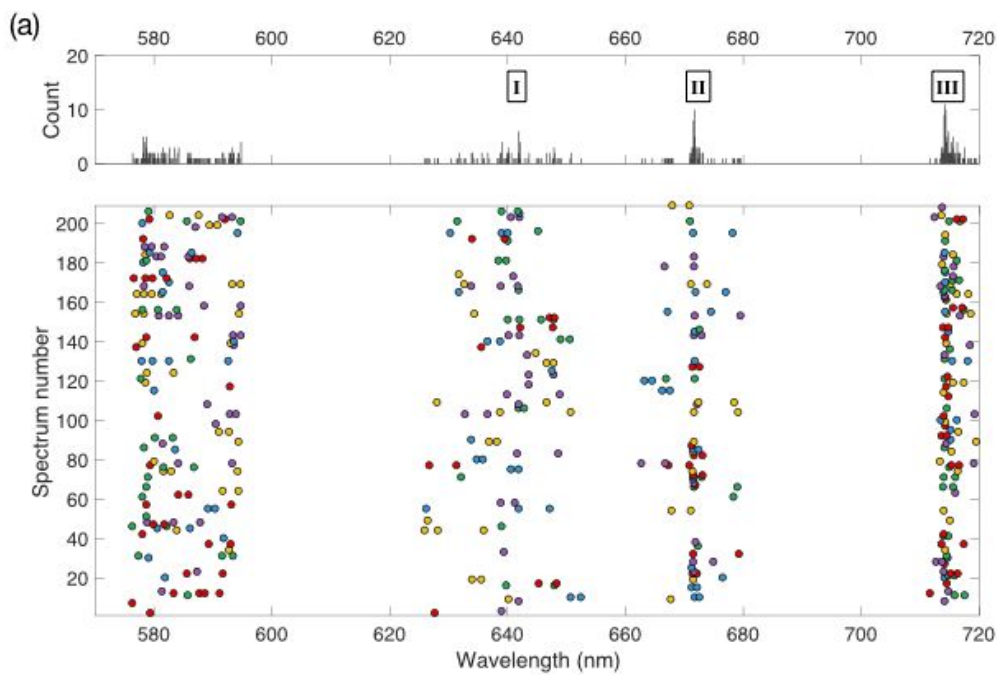

*Figure S5: Cryogenic emission under 532 nm excitation. (a) Histogram of intensity peaks sampled over 205 pillars. Circles indicate the center of an observed emission peak, and colors correspond to individual pillars horizontally across spectral regions. (b) Comparison between conditional probabilities and individual probabilities for the observation of emission in each region.*

**Photostability of Pb Emitters**

Some Pb emitters show instability in their fluorescence emission over time. Figure S6 shows spectra of an emitter switching from a bright state, in which emission at 515 and 520 nm is present, and a dark state in which no emission is observed within that region. Interestingly, the emission wavelength of the line near 640/670 nm is correlated to the switching of the 520 nm line. This could imply a changing strain environment as the Pb emitter changes charge states via photoionization, similar to the SiV or NV, or as an indication that emission around 640 nm is also Pb-related. Further investigation is required to determine the mechanism of the shift in emission frequency. In addition to the bistability shown above, other Pb emitters show photobleaching without recovery under continuous 450 nm illumination.

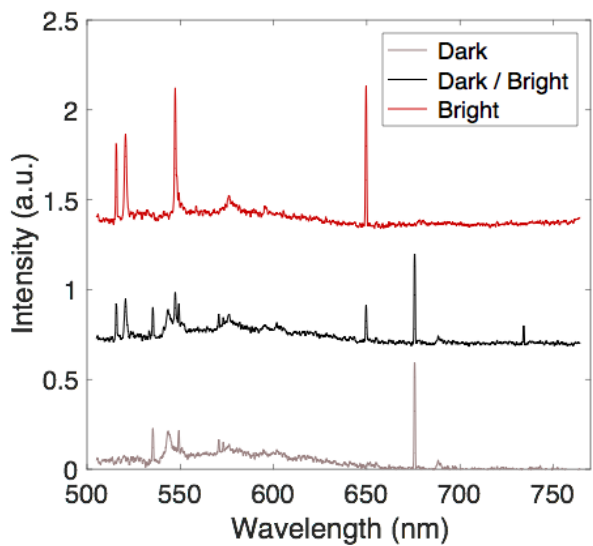

*Figure S6: Photo-switching of a Pb-related emitter. Temporally separated spectra (~ 1 min) show distinct emission profiles.*

**Polarization and Temperature Studies**

We present the polarization and temperature dependence of the representative emitter from Figure 4b (reproduced and color coded in Figure S7). The polarization of the pair of lines at 515 nm (red) and 520 nm (black) are nearly orthogonal as indicated in Figure 5b. All but one line (blue) in Region II (545-550 nm) appear to be similarly polarized. In Figure 5c, we observe that the line at 515 nm broadens from a spectrometer-limited linewidth of 0.1 nm (120 GHz) at 4K to 0.18 nm (200 GHz) at 100K and 0.86 nm (970 GHz) at 160K, whereas the line at 520 nm has a linewidth of 0.99 nm (1100 GHz) even at 4K. The relative fluorescence intensity also changes significantly with temperature, possibly indicating that these lines correspond to two excited state transitions to a single ground state as with other group IV emitters.

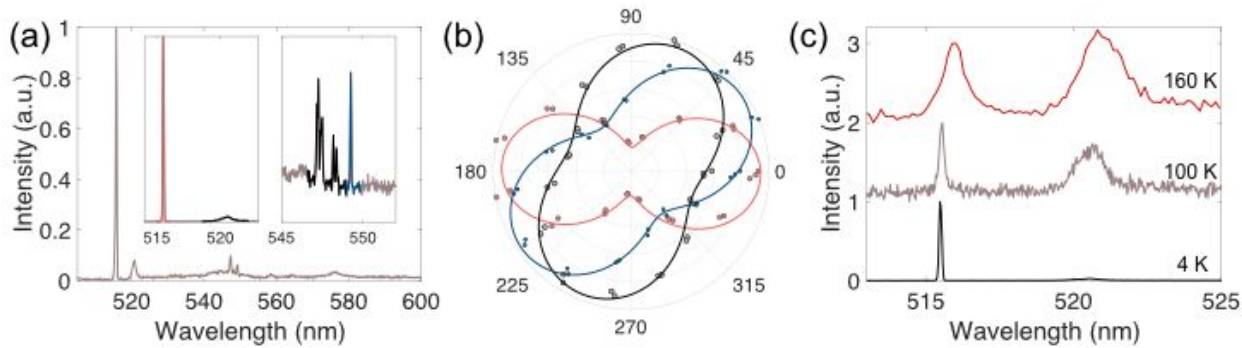

*Figure S7: Polarization and temperature dependence. (a) Zoom-in of emission spectrum from boxed emitter in Fig. 3a,b. Inset: Region I and Region II. (b). Polarization analysis of Pb spectral lines at 515 nm (red) and 520 and 547 nm (black), and 548 nm (blue). (c) Temperature dependence of Pb-emitter fluorescence, Region I.*

**Acknowledgements**

M.T. acknowledges support by an appointment to the Intelligence Community Postdoctoral Research Fellowship Program at MIT, administered by Oak Ridge Institute for Science and Education through an interagency agreement between the U.S. Department of Energy and the Office of the Director of National Intelligence. N.H.W is supported in part by the Army Research Laboratory Center for Distributed Quantum Information (CDQI). E.B. was supported by a NASA Space Technology Research Fellowship. D.E. and experiments were supported in part by the STC Center for Integrated Quantum Materials (CIQM), NSF Grant No. DMR-1231319. K.C.C. acknowledges funding support by the National Science Foundation Graduate Research Fellowships Program (GRFP). This research used resources of the National Energy Research Scientific Computing Center, a DOE Office of Science User Facility supported by the Office of Science of the U.S. Department of Energy, as well as resources at the Research Computing Group at Harvard University.